\documentclass[11pt]{iopart}
\pdfoutput=1
\usepackage{multirow} 
\usepackage[figuresright]{rotating}

\begin{document}
\title{The GAP-TPC} 


\author{B.~Rossi$^{c,b}$, A. Anastasio$^b$, A. Boiano$^b$, S. Catalanotti$^{a,b}$, A. G. Cocco$^b$, G. Covone$^{a,b}$, P. Di Meo$^b$, G. Longo$^{a,b}$, A. Vanzanella$^b$, S. Walker$^{a,b}$, H. Wang$^d$, Y. Wang$^d$, G. Fiorillo$^{a,b}$\\}
\address{$^a$ Universit\`a degli Studi di Napoli "Federico II", Via Cinthia, I-80125 Napoli, Italy }
\address{$^b$ INFN, Sezione di Napoli, Via Cinthia, I-80125 Napoli, Italy}
\address{$^c$ Department of Physics, Princeton University, Princeton, New Jersey 08544, USA }
\address{$^d$ Department of Physics and Astronomy, University of California, Los Angeles, CA 90095, USA}

\ead{rossib@princeton.edu}

\begin{abstract}
Several experiments have been conducted worldwide, with the goal of observing low-energy nuclear recoils induced by WIMPs scattering off target nuclei in ultra-sensitive, low-background detectors.  In the last few decades noble liquid detectors designed to search for dark matter in the form of WIMPs have been extremely successful in improving their sensitivities and setting the best limits. One of the crucial problems to be faced for the development of large size (multi ton-scale) liquid argon experiments is the lack of reliable and low background cryogenic PMTs: their intrinsic radioactivity, cost, and borderline performance at 87 K rule them out as a possible candidate for photosensors. We propose a brand new concept of liquid argon-based detector for direct dark matter search: the Geiger-mode Avalanche Photodiode Time Projection Chamber (GAP-TPC) optimized in terms of residual radioactivity of the photosensors, energy and spatial resolution, light and charge collection efficiency
\end{abstract}

\maketitle
\section{Introduction}
%
\makeatletter
\@addtoreset{footnote}{section}
\makeatother

Current dark matter detectors using noble liquids have an effective target mass ranging from 100 kg to the ton-scale (e.g. LUX~\cite{[LUX]}, Xenon-1T ~\cite{[Xenon]}, DarkSide ~\cite{[DS50b],[DS50a]}). Hundreds of 2-3 inch photomultiplier tubes\footnote{Hamamatsu PMTs' R11065 and R11410-20 are respectively used in Darkside-50 and Xenon-1T experiments, while R8778 PMTs are used in LUX experiment.} (PMTs) are used in these detectors for the accurate measurement of scintillation light from liquid argon (128 nm shifted to 420 nm) and liquid xenon (170 nm). A prerequisite for the next generation multi-ton scale noble liquid experiments to reach the so-called {\it neutrino floor}\footnote{Sensitivities for the WIMP nucleon scattering cross section of $\sigma\sim$10$^{-49}$ cm$^2$ at 100~GeV/c$^2$.} is the reduction of radiogenic background sources to negligible levels. Photomultipliers as scintillation light readout devices and cryostat shells and flanges represent the largest sources of radioactive background for current experiments.

While the radioactive background budget is the primary issue, it is not the only difficulty inherent in using PMT technology as it stands today. In particular for the liquid argon case, current cryogenic PMTs (R11065 and R10065-20) do not fully match the requirements of future experiments in terms of performance.

We propose a brand new concept of detector for dark matter search: the Geiger-mode Avalanche Photodiode Time Projection Chamber {\bf (GAP-TPC)}. The GAP-TPC is designed to achieve: a very high light collection, an exceptional single photon resolution and a sub-cm spatial resolution on the XY plane that will considerably improve the surface background rejection. 

The larger photosensitive coverage of the detector surfaces needed to reach the high design light yield requires a significant increase of the number of photodetectors. Our idea is to use SiPM arrays that are virtually radioactivity free, given the reduced mass and provided they are packaged with radiopure materials. In addition, they allow for cost effective large productions. The use of SiPMs reduces the complexity of the detector because they need a common low bias voltage to operate, with no need for high voltage feedthroughs and cables. Moreover, SiPM compactness reduces to the minimum the amount of passive liquid argon (LAr) around the detector and opens up the possibility of increasing the photocathodic coverage of the TPC. Given the above considerations, we propose a ${\bf 4\pi}$ design to directly convey the photons to the photosensors with the highest efficiency without any use of reflectors. 
%
\section {The GAP-TPC: Concept detector structure}
\label{sec:structure}
\indent
%

\makeatletter
\@addtoreset{footnote}{section}
\makeatother

\begin{figure}[!htp]
\center\includegraphics[width=0.7\linewidth]{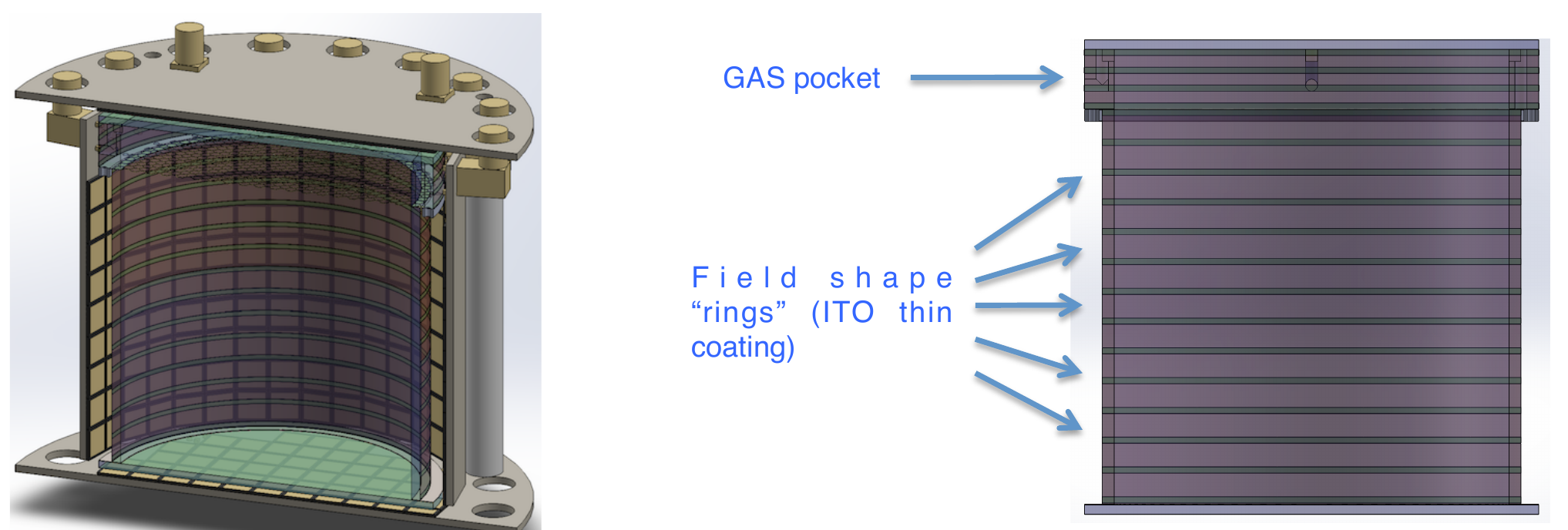}
\caption{Left: Conceptual CAD sketch of the GAP-TPC. Right: The GAP-TPC active volume is defined by the fused silica vessel. An ITO coated-fused silica window in the bottom part of the cylinder acts as TPC cathode. On the top part, a grid extracts ionization electrons to the gas pocket and a second ITO coated-fused silica window provides the TPC anode. The drift field is kept uniform along the drift coordinate by means of ITO field shapers deposited on the walls of the fused silica vessel.}  
\label{fig:GAP-TPC_detector}
\end{figure}

The GAP-TPC is a ${\bf 4\pi}$ photocatodic coverage liquid argon Time Projection Chamber detector that uses SiPM arrays (see fig.~\ref{fig:GAP-TPC_detector}). The ${\bf 4\pi}$ design allows for the direct convey of the photons to the photosensors with the highest efficiency without any use of reflectors and for maximizing the photocathodic coverage. 
A fused silica vessel defines the active volume. To the bottom part of the cylinder is placed the cathode, an ITO coated-fused silica window. On the top part, there is the grid for extracting ionization electrons to the gas pocket, the ``diving bell'' for creating the gas region as well as a second ITO coated-fused silica window acting as an anode. The drift field is applied between the cathode and the anode and is kept uniform along the drift coordinate by means of field shapers. In conventional LAr TPCs the shapers are metallic rings encapsulating the drift region. This solution is not applicable to a $4\pi$ coverage. Our idea is to realize the field shaping rings by thin coating the walls of the fused silica vessel with ITO (fig.~\ref{fig:GAP-TPC_detector}). Finally, a wavelength shifter (TPB) is carefully deposited on the vessel and planar surfaces to convert the Ar 128~nm scintillation light to 420~nm. The ITO very high transparency to visible light, together with the $4\pi$ optical readout and the overall larger QE of the SiPM ($>$50\%)\footnote{http://sensl.com/products/j-series/}  with respect to cryogenic PMTs (<40\%) allows for the expected significant increase in the total light collection. 
There are conceivable implementations of this design that can be scaled up to very large volumes.

\begin{figure}[t]
\center\includegraphics[width=0.7\linewidth]{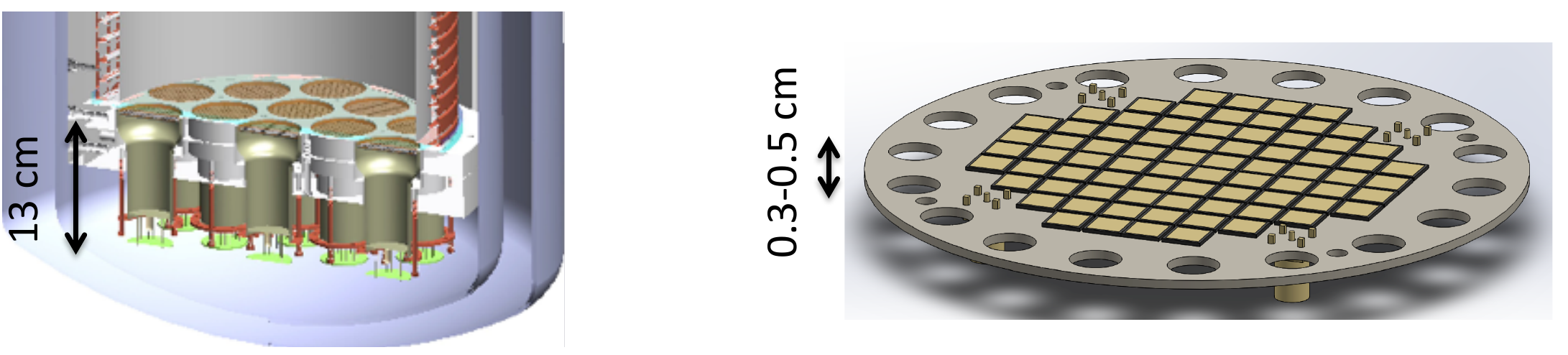}
\caption{Sketch comparing the current LAr TPC configuration with PMTs (left) and the innovative design of the GAP-TPC that make use of thin arrays of SiPMs (right). Thanks to the smallness of the SiPM arrays, the GAP-TPC will minimize the passive layer of cryogenic liquid surrounding the detector.} 
\label{fig:top_array}
\end{figure}
An example of the bottom photosensor plane of a GAP-TPC prototype with an internal cylindrical shape is shown in fig.~\ref{fig:top_array}. The compactness of the SiPMs with respect to PMTs reduces to minimum the amount of passive LAr around the detector increasing the ratio of active/not-active Argon contained in the cryostat. 
The top and bottom photosensor planes can be composed of different SiPM array units. For the lateral coverage of a cylindrical region one can foresee vertical slim arrays to be installed all around the fused silica cylinder to achieve the $4\pi$ coverage.  SiPMs give an additional advantage. The readout pattern of the photosensors' top array can be chosen depending on the needs. As example, SiPMs can be readout with higher granularity nearby the edges of the TPC where the XY resolution is key to enhance surface background rejection and with much less granularity in the center of the top plane where high resolution on the position reconstruction is not needed. 
%
%

It is not trivial to sum up the output of SiPM dies. A possibility is the analog sum of multiple SiPM dies (e.g. summing up 64 SiPM output and reading out with a single cable) by means of a cryogenic transimpedance amplifier. Important results on this matter have been already published in our previous work~\cite{[ArrayB-32]}. 
%
\section {GAP-TPC prototype: Experimental setup}
\label{sec:exp_setup}
\indent

\begin{figure}[!ht]
\center\includegraphics[height=5cm]{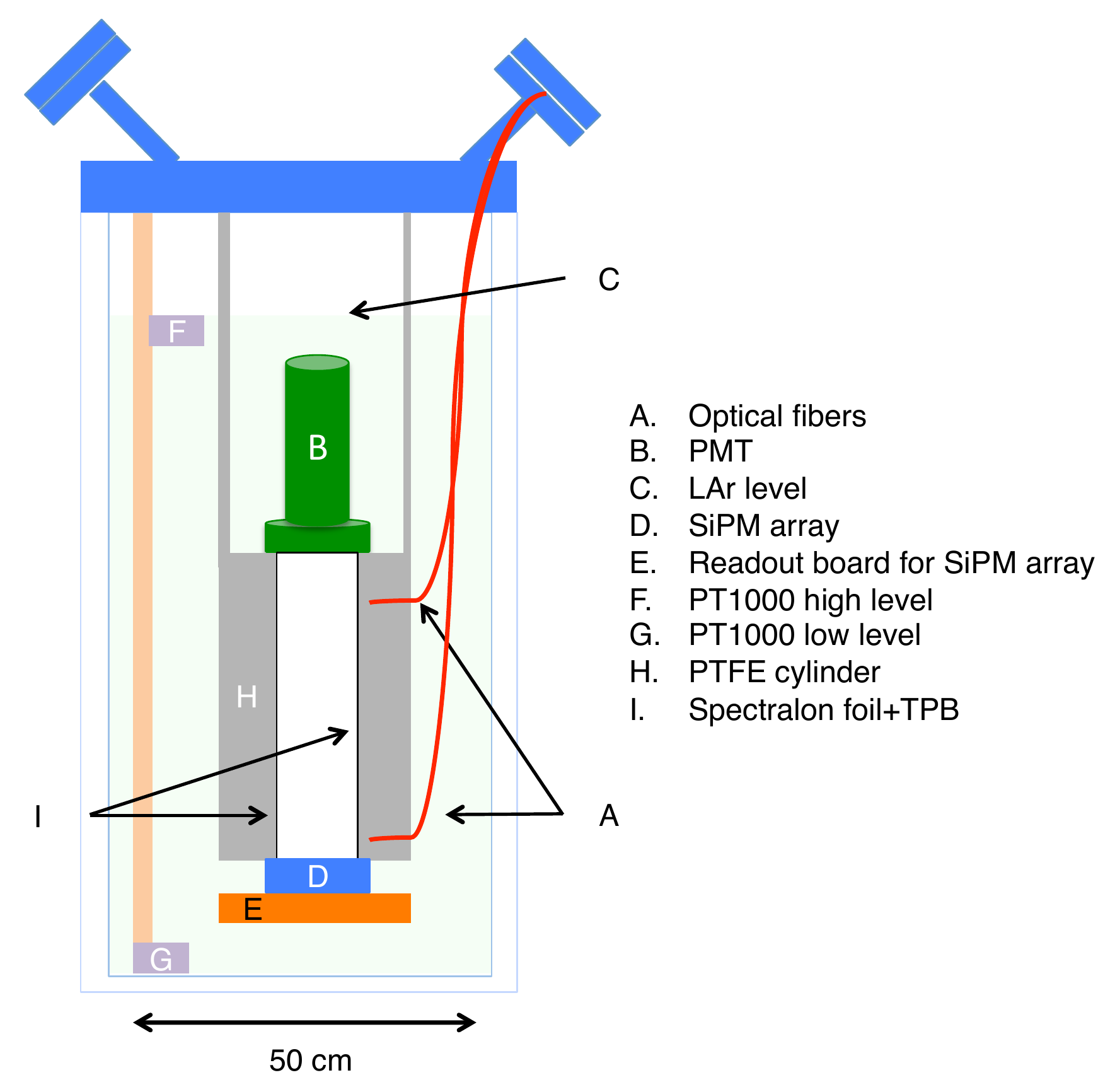}
\caption{Sketch of the experimental setup.} 
\label{fig:gap_prototype}
\end{figure}

An embryonal GAP-TPC has been equipped at the INFN Napoli cryogenic laboratories to compare the performance at the liquid Argon temperature of a cryogenic Hamamatsu 3~inch R11065 PMT and a 6$\times$6~cm$^2$ SensL-ArrayC-60035-64P\footnote{http://sensl.com/products/c-series} SiPM array. The LAr detector prototype has an active volume bounded by a PTFE cylinder of 100~mm height and 57~mm internal diameter.  An enhanced specular reflector (Lumirror foils) was coated with the TPB wavelength shifter in use for Darkside-50 experiment~\cite{[DS50b]} and rested against the internal wall of the PTFE cylinder. The detector was equipped with the two photosensors: the PMT installed on the top of the PTFE cylinder and the SiPM array installed on the bottom of the cylinder (fig.~\ref{fig:gap_prototype}). 

The SiPM array is composed of 64 individual SiPMs of 6.0$\times$6.0~mm$^2$ active area (7.2$\times$7.2~mm$^2$ total surface area),  each containing 18980 $\mu$-cells of 35$\times$35~$\mu$m amounting to a total active area of 57.2$\times$57.2~mm$^2$.   Both photosensors were coated with uniform and similar layers of TPB.  The choice of the PTFE internal diameter was made to ensure that the openings at both ends were fully covered by the photosensing area, in order to ease the comparison of the individual light yields.
%
%

The detector is housed in a double wall stainless steel cryostats that contain the cryogenic liquid. A top flange vacuum-seals the cryostat through an indium wire. The flange hosts a series of additional smaller flanges and feedthroughs needed to evacuate the inner volume of the vessel, to fill in the liquid, to read-out the detector signals, to supply the bias voltage, to house the pressure monitors, level meter and safety valves, and to provide the transport of laser beams to the inner part of the TPC, used as a calibration and monitoring tool for the photosensors.

Light pulses are generated by a Hamamatsu PLP10 light pulser, equipped with a 408~nm laser head with a pulse width of 70~ps. The laser beam is passed through an optical attenuator mount in which discrete filters of different attenuation coefficients were used to select the magnitude of illumination reaching the photosensors. After the attenuator, a custom made 1x2 fiber splitter brings the laser light to cryostat. Two internal fibers are installed at different heights along the vertical direction of the FTPE cylinder, one nearby the PMT and the other one nearby the SiPM array, to ensure both photosensors to be reached by the laser light. 

Before the data taking starts, the cryostat is pumped out until a residual pressure of 10$^{-4}$~mbar is reached. Then, the LAr filling operation starts. Commercial argon gas is flown through a commercial getter (OXYSORB) in order to remove impurities (mainly oxygen, nitrogen and water) before to be inserted in the cryostat. The level of the liquid argon in the cryostat is monitored with two PT1000 resistors, readout by a calibrated CRYOCON 32 controller. The first PT1000 (low level) is placed below the SiPM array readout electronic board, while the second one (high level) is placed a few cm above the PMT base. The liquid argon filling operation is stopped 15~min after the high level is reached. The LAr level is continuously monitored in order to ensure constant thermodynamical conditions during the data taking. 

The bias voltage was supplied via a low noise power supply\footnote{TTi QL355TP} through a 10~k$\Omega$ resistor. The SiPM array was coupled to a custom made cryogenic board able to sum up 32+32 SiPM outputs with only two channels, developed at the electronic workshop of INFN Napoli. 
The two outputs of the readout board are conveyed outside the cryostat through a signal feedthrough flange and are fed to an external (custom made)\footnote{At the electronic workshop of LNGS} NIM amplifier (gain x10). The output of the amplifier is fed in to a CAEN V1720E digitizer with 4~ns sampling and 12~bit resolution, connected via optical link to a PC for data handling, storage and analysis.
%
\subsection {Calibration through gamma source}
\indent

\begin{figure}[t]
\center\includegraphics[width=1.\linewidth]{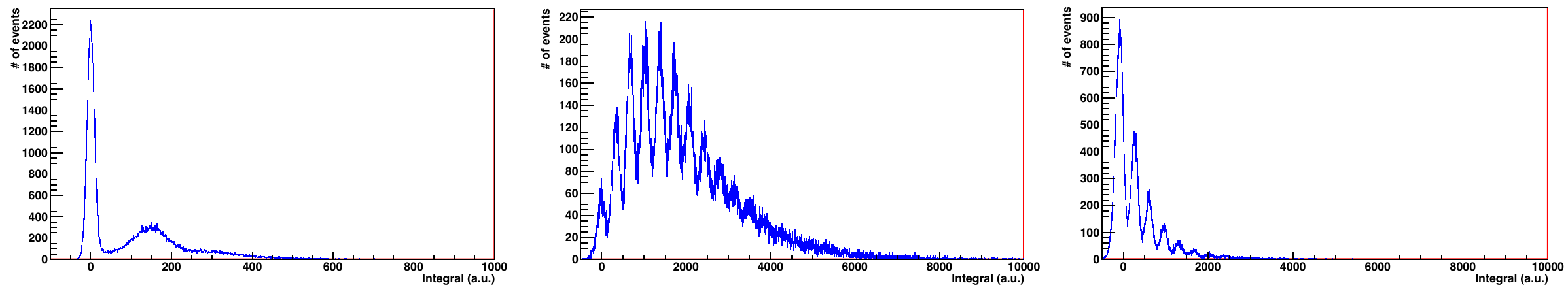}
\caption{a) Single photon spectrum of the Hamamatsu R11065 PMT. b-c) SPE spectum of the two outputs of the cryogenic readout board of the SensL-ArrayC-60035-64P.  The SiPM array photon counting capabilities are clearly excellent.} 
\label{fig:SPE}
\end{figure}

The single photoelectron (SPE) response of the photosensors have been measured and monitored during the data taking. Each data taking run consisted of 100,000 triggers at a given V$_\mathrm{bias}$, with a memory buffer of 5~$\mu$s, with 2~$\mu$s pre-trigger. The laser was triggered by an external pulser at a repetition rate of 10~kHz and its illumination was set, through a system of discrete attenuators in order to send a few photons for each trigger to the photosensor. The single photoelectron spectrum (fig.~\ref{fig:SPE}), was reconstructed integrating the waveform region around the trigger for 800~ns\footnote{The integration window was set as large as to contain at least three time the recharge time (3$\tau$).} for a given run. 
In addition, a measurement of the influence of cross-talk and afterpulsing on photodetection characteristics has been performed on the SiPM array. The method used is described in~\cite{[Vinogradov]}. 

During the data taking, the operational voltage for the PMT was set to 1250~V, while the operational voltage for the SiPM array was set to V$_\mathrm{OV}$=6.3~V (overvoltage), where the probability of correlated pulses were evaluated to be $\sim$25\%.

\begin{figure}[!ht]
\center\includegraphics[height=4cm]{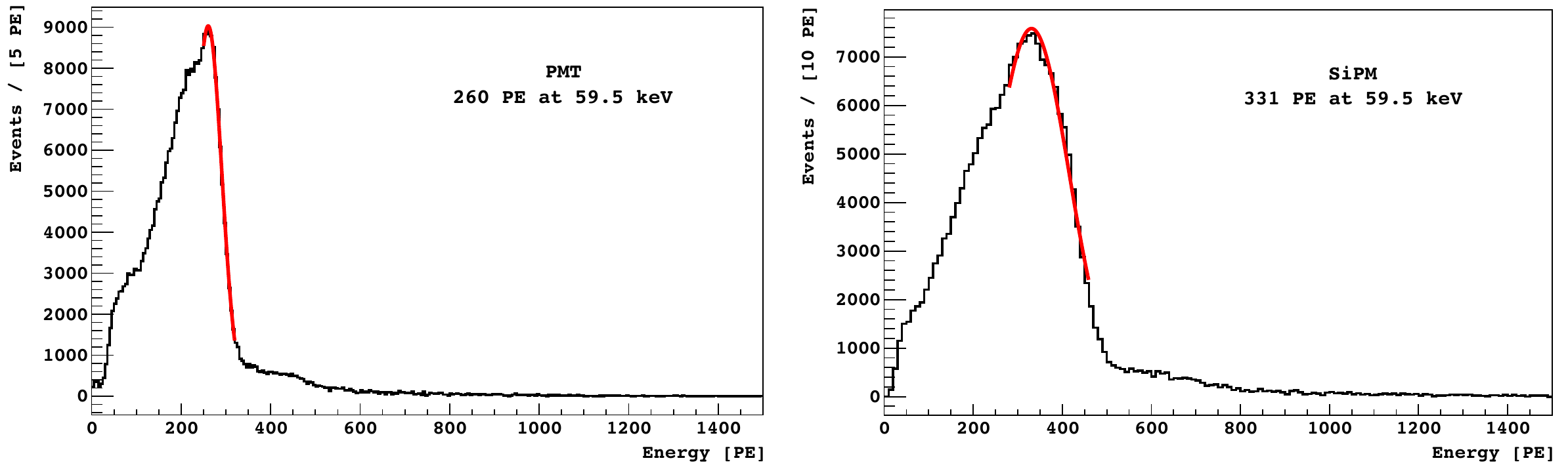}
\caption{Tha gamma-ray Americium peak reconstructed with the Hamamatsu PMT (left) and with the SiPM array (right). The light yield value of the SiPM is already rescaled to take into account the effect of correlated pulses (optical cross talk and afterpulses). The GAP-TPC prototype shows a light yield of 9.9~PE/keV, the maximum ever reached for a liquid argon detector. The SensL SiPM array exceeded the light yield of the Hamamatsu PMTs by more than 20\%.}
\label{fig:LY}
\end{figure}

The GAP-TPC prototype was calibrated with the use of an Am-241 gamma-ray source (59.5~keV). The source was placed outside the cryostat in a lead collimator of cylindrical size of 10~mm diameter and 30~mm length, pointing towards the geometrical center of the active detector, at the same distance between the PMT and the SiPM array. The overall light yield was found to be LY=9.9~PE/keV. The PMT light yield was LY=4.4~PE/keV while the SiPM array achieved Y=5.5~PE/keV\footnote{The light yield value of the SiPM array has been rescaled to take into account the effect of correlated pulses.} (fig.~\ref{fig:LY}). The measured value of LY of the GAP-TPC prototype set a record for liquid argon detectors, being 20\% higher of what achieved so far by the DarkSide-50 detector. Furthermore, with this test we have also demonstrated that the SensL SiPM array has a light yield about 20\% greater than the PMT. SiPM technology is evolving very fast, so it is very reasonable to predict an increase of their QE in the near future. 
%
\section {Conclusions}

SiPMs look to be very promising devices for next generation noble liquids direct dark matter search experiments. SiPM arrays sizes are nowadays comparable to PMTs of 3 inch size. The manufacture progress of last years have made SiPMs arrays very appealing for substituting PMTs in cryogenic environment. In particular, SensL-ArrayC-60035-64P can be operated at liquid argon temperature coupled with a cryogenic readout amplifying board, to reduce the number of output channels without distorting the pulse shape. We assembled and operated in liquid argon the GAP-TPC prototype. The calibration of the detector resulted in a record light yield of LY=9.9~PE/keV. The SiPM array showed much better performance with respect to the PMT in terms of light collection. 

Light collection is the key figure of merit for a LArTPC, driving the numerical performance of the strong intrinsic pulse shape discrimination of LAr.  These results represent a very important step towards the adoption of SiPMs as the baseline for future liquid argon dark matter detectors.
%
%
%
\section* {Acknowledgements}
%

The work presented in this paper was conducted thanks to grant PHY-1314507 National Science Foundation and from INFN. We warmly acknowledge both institutions.


\section*{References}
\begin{thebibliography}{9}
\bibitem{[LUX]} D.S. Akerib et al., "First Results from the LUX Dark Matter Experiment at the Sanford Underground Research Facility", Phys. Rev. Lett. 112 (2014) 091303
\bibitem{[Xenon]} E. Aprile et al., "Physics reach of the XENON1T dark matter experiment", arXiv:1512.07501.
\bibitem{[DS50b]} P. Agnes et al.,  "First Results from the DarkSide-50 Dark Matter Experiment at Laboratori Nazionali del Gran Sasso", Phys.Lett. B 743 (2015) 456-466
\bibitem{[DS50a]} C.E. Aalseth et al.,  "The DarkSide Multiton Detector for the Direct Dark Matter Search", Adv.High Energy Phys. (2015) 541362
\bibitem{[Vinogradov]} S. Vinogradov et al., {\emph 2009 IEEE Nuclear Science Symposium Conference Record}, N25-111 (2009)
\bibitem{[ArrayB-32]} S. Catalanotti et al., {\emph Performance of a SensL-30035-16P Silicon Photomultiplier array at liquid argon temperature}, JINST 10 (2015) P08013. 
\end{thebibliography}
\end{document}